\documentstyle[12pt]{article} \hoffset-1.5cm
\begin{document} \begin{center}

{\huge Hadronic $\gamma$-ray emission models} 
\vskip.5cm
 
{\it Karl Mannheim (kmannhe@uni-sw-gwdg.de)}
\vskip.5cm
\end{center}

Models for $\gamma$-ray emission from active galactic
nuclei (AGN) based on hadronic
interactions have in the past considered interactions 
between relativistic protons and
matter in an accretion disk surrounding a black hole.
However, it has become evident that 
$\gamma$-rays originate in relativistic jets in which the matter density
is too low to provide the required target matter for relativistic protons.  
The density of photons, on the other hand,
is large enough to cool relativistic
protons efficiently provided their energy exceeds the threshold for secondary
particle production (pairs, pions, kaons, ...).  
In spite of the hadronic cross-section $\sigma_{\rm p\gamma}=5\, 10^{-28}$~
cm$^2$ being much smaller than the Thomson cross-section $\sigma_{\rm T}=
6.7\, 10^{-25}$~cm$^2$, the proton cooling time scale can be as
short as the electron cooling time scale.
Comparing pion photo-production by protons with
inverse-Compton scattering of electrons on the same photon target 
with photon density $n_\gamma\propto \epsilon^{-1}$, 
one obtains
\begin{equation}
{t_{\rm p}\over t_{\rm e}}\simeq {n_\gamma(\epsilon){\epsilon\over m_{\rm e}c^2}
\gamma_{\rm e}\sigma_{\rm T} c\over
n_\gamma({m_\pi c^2\over \gamma_{\rm p}}){m_\pi\over m_{\rm p}}\sigma_{\rm p
\gamma} c}={m_{\rm p}\over m_{\rm e}}{\gamma_{\rm e}\over \gamma_{\rm p}}
{\sigma_{\rm T}\over \sigma_{\rm p\gamma}}\approx 2.5\, 10^6{\gamma_{\rm e}
\over \gamma_{\rm p}}\ ,
\end{equation}
i.e. $t_{\rm p}=t_{\rm e}$ for $\gamma_{\rm p}=2.5\, 10^6\gamma_{\rm e}$.
Consequently, the proton-induced luminosity given by
\begin{equation}
L_{\rm p}\simeq {u_{\rm p}\over u_{\rm e}}{t_{\rm e}\over t_{\rm p}}
L_{\rm e}
\end{equation}
can exceed the Compton
luminosity by a large factor (e.g., the relativistic proton-to-electron
energy density ratio 
in the Galaxy is $u_{\rm p}\approx 100 u_{\rm e}$).
In a statistical acceleration process such as Fermi acceleration at shock
waves, acceleration operates until energy losses (or geometrical constraints)
become important implying much larger proton than electron maximum energies.  
Photoproduction by shock-accelerated protons in a
relativistic jet at some distance to the central black hole is assumed in the
models of Protheroe (1996a,b) and Mannheim (1993a).\\

  These models assume
different target radiation spectra.  Protheroe assumes a disk spectrum, whereas
a synchrotron spectrum is assumed in the latter model.  The disk spectrum is
expected to be more important if particle acceleration occurs at distance less
than a parsec from the central black hole and the synchrotron spectrum dominates
at larger distances.  If there is continuous proton acceleration along the jet,
both targets fields are important.  Shock acceleration of particles outside of
the central parsec is expected to accompany the passage of the supersonic jet
through the steep external pressure gradient characterizing the external medium
beyond the Broad Line Region.  Some jet formation models predict magnetic
collimation to a cylindrical flow with a transverse
radius of the order of 
$r_{\rm j}\sim 100 r_{\rm G} = 3\, 10^{15}m_8$~cm.  If the collimation
can be maintained over a parsec,
rapid variability is possible even in the synchrotron target model
on time scales of the order of
$\sim r_{\rm j}/(\delta_{\rm j}c) \sim 10^4 \delta_{10}m_8 $~s 
where $m_8$ denotes black hole mass in units of $10^8M_\odot$ and 
$\delta_{\rm j}$ is the jet Doppler factor
in units of 10.\\

Both models find that the $\gamma$-ray spectra from hadronic synchrotron cascades
can explain the observed properties of $\gamma$-ray emitting AGN.  They also
predict very similar diffuse neutrino fluxes from the sources.  If $\gamma$-ray
emitting AGN produce the isotropic diffuse $\gamma$-ray background, the
corresponding neutrino background at high energies can be detected by a
kubic-kilometer underice muon detector.  The models further agree that 
(comoving frame) proton
energies in the range $10^8-10^{10}$ GeV are required to explain the
observations.  At perpendicular relativistic shocks, the maximum
rate of energy gain is $\dot{E} \sim 0.4 e c^2 B$ 
(corresponding to acceleration at almost the gyro-time).
This is sufficient to achieve the necessary high energies in a time
interval shorter than the (comoving frame) variability time scale.\\

The discovery of TeV emission from nearby BL Lacertae objects entails an
important implication regarding the distance of the $\gamma$-ray emission zone from
the central accretion flow.  In the framework of a unified model for AGN, in
which the difference between moderate-luminosity radio galaxies and BL
Lacertae objects is due to different orientations of jet and circum-nuclear dust torus
with respect to the observer, a strong central near-infrared radiation field
is expected from the warm ($\sim$~1000~K) inner edge of the dust torus with a
thickness of a few parsecs.  Owing to this warm dust torus, 
TeV $\gamma$-rays emitted
from within the central parsec (predicted in external photon
inverse-Compton models) are converted into pairs and
degraded to lower energies (Mannheim 1993b, Protheroe \&
Biermann 1996). 
Furthermore, the rapid variability of the TeV $\gamma$-ray emission from Mrk421 on
a time scale of hours and its correlation with emission at lower photon energies (see contribution by M.~Dietrich, this volume)
contradict the predictions of the Blandford \& Levinson
(1995) pair-induced cascade model in which the size of the $\gamma$-ray emission zone
increases with photon energy.\\

Thus, it appears that the hadronic models compete with
synchrotron-self-Compton (SSC) models in explaining the TeV $\gamma$-ray emission
from nearby blazars.  Fitting the multifrequency spectrum of Mrk421 with either
model yields very different values of the magnetic field strength.
Whereas Protheroe (1996a) and Mannheim (1996) find values of 30 and 40
G, respectively,
the SSC model of Stecker et al.  (1996) yields 0.2 G.  The  
difference by a factor of 200 has important physical implications. 
The apparent nonthermal luminosity of Mrk421 
in bright states is $\sim 5\, 10^{45}$~ergs~s$^{-1}$
in the UV-to-soft--X-ray band which corresponds to an emitted luminosity
of the order of $\sim 10^{42}\delta_{10}^{-4}$~ergs~s$^{-1}$ owing to Doppler boosting
with $\delta_{10}\sim \delta/10$.  
In order to produce the observed emission, 
the kinetic luminosity of the jet in Mrk421 must be at least  
$\sim 10^{43}$~ergs~s$^{-1}$ (assuming a radiative efficiency of 10\%).
Hence it follows that the SSC (comoving frame)
magnetic field strength of 0.2~G at $r\sim 3\, 10^{15}$~cm
is insufficient
to confine the jet. On the
other hand, to keep the jet radius small even at a distance of more than one
parsec away from the black hole magnetic jet collimation is required.  External
pressure confinement would overproduce X-rays by its free-free
emission and the half-opening
angle $\phi\propto 1/\gamma_{\rm j}$ of a free jet would 
correspond to a jet Lorentz
factor of $\gamma_{\rm j}> 10^3$.  A jet with a Lorentz factor larger
than $\sim 10$ cannot
escape the central parsec owing to the Compton drag.
Another argument against low magnetic field values at the milliparsec 
transverse scale is
that the jets in Fanaroff-Riley type II galaxies
with kinetic luminosities of $\sim 10^{46}$~ergs~s$^{-1}$
still have magnetic field strengths of 100-300 $\mu$G 
at the kiloparsec transverse scale (Meisenheimer et al.~1989,
Harris et al.  1994).  An adiabatic compression of these fields 
($B_\perp\propto r^{-1}$)
yields $\sim 100-300$ G at the milliparsec scale.  Scaling with
the jet luminosity $B\propto \sqrt{L}$ then yields a (comoving frame) 
field strength of the order of
$\sim 10$~G for the $\gamma$-ray emitting zone in the jet of Mrk421.
It is obvious that a
detection of blazars at energies above TeV would require still lower
magnetic field strengths in the SSC models which would make the problem 
even more
severe than it already is.  \\

The hadronic emission models predict emission  above TeV
(e.g., Mannheim et al.~1996) and if it were absent, this
would rule out the hadronic emission models.  Cosmic absorption of
$\gamma$-rays by pair production in collisions with low-energy
diffuse background photons, however, represents a major obstacle for this
crucial experiment. Nevertheless,
a surprisingly strong $\gamma$-ray flux has been tentatively detected 
with HEGRA at 50 TeV by co-adding
the events from the positions of the nearest blazars for which cosmic
absorption by collisions with diffuse $\sim 100\mu$m
photons is expected to be lowest (see contribution
by H.~Meyer, this volume).
According
to the hadronic emission models, electromagnetic 
power injected at much higher energies than TeV is reprocessed
by a synchrotron cascade towards lower energies. It depends
on the pair creation opacity of the emission region 
for $\gamma+\gamma\rightarrow e^++e^-$
below which energy the reprocessed
photons emerge.  In the proton blazar model (Mannheim 1993a), 
synchrotron emission by accelerated electrons is calculated for
all jet radii from $r_{\rm ir}$ to some $r_{\rm r}>r_{\rm ir}$, 
i.e. from the radius where the infrared photons
become optically thin to the much larger radius where the
low-frequency radio photons are produced.
Cascade radiation was considered only from the radius $r_{\rm ir}$,
since the infrared photons are the most important target photons
for the accelerated protons.
At $r_{\rm ir}$, the turnover energy is $\sim$ TeV and the cascade
spectrum above TeV should steepen by one power compared with the sub-TeV
spectrum owing to the energy-dependent
pair-creation opacity.  This is in accord with recent measurements of Mrk421 with the
HEGRA air-$\check{\rm C}$erenkov telescopes (Petry et al.~1996)
and justifies the assumption that the cascade emission is dominated by
emission from $r_{\rm ir}$.  However, cascade emission above
TeV from an inhomogeneous jet
at $r>r_{\rm ir}$ is still possible and could produce additional
$\gamma$-ray components.
Reliable estimates of the
$\gamma$-ray flux above TeV  require further calculations of
proton-initiated cascades for the entire jet.
The emission above TeV would then be less variable than
the emission at TeV -- analogous to the effect of
an increasing $\gamma$-ray
photosphere invoked in the pair-induced
cascade model of Blandford \& Levinson (1995), albeit at higher
energies. 

\vskip0.5cm

\noindent
{\bf References: }
\vskip0.5cm

\noindent\hangindent=0.8cm 
Blandford, R., Levinson, A., 1995, ApJ 441, 79

\noindent\hangindent=0.8cm 
Harris, D.E., et al., 1994, Nature 367, 713 

\noindent\hangindent=0.8cm 
Mannheim, K., 1993a, A\&A 269, 67 

\noindent\hangindent=0.8cm 
Mannheim, K., 1993b, Phys.Rev.D 48, 5270 

\noindent\hangindent=0.8cm 
Mannheim, K., 1996, Sp.Sc.Rev. 75, 331 

\noindent\hangindent=0.8cm 
Mannheim, K., Westerhoff, S., Meyer, H., Fink, H.H., 1996, A\&A
315, 77 

\noindent\hangindent=0.8cm 
Meisenheimer, K., et al., 1989, A\&A 219, 63

\noindent\hangindent=0.8cm 
Petry, D., \& HEGRA collaboration, 1996, Astropart.~Phys. 4, 199

\noindent\hangindent=0.8cm 
Protheroe, R.J., 1996a, submitted to Astropart. Phys.
 
\noindent\hangindent=0.8cm 
Protheroe, R.J., 1996b, to appear in Accretion Phenomena and Related
Outflows, IAU Colloq.~163, ed. D.~Wickramashinghe et al.
(astro-ph/9607165)

\noindent\hangindent=0.8cm 
Protheroe, R.J., Biermann, P.L., 1996, 
accepted for publication in Astropart. Phys. (astro-ph/9608052) 

\noindent\hangindent=0.8cm 
Stecker, F.W., et al., 1996, submitted to ApJ Letters
(astro-ph/9609102) 

\end{document}